\documentclass[aps,twocolumn,pra,tightenlines,floatfix,superscriptaddress]{revtex4-2}

\usepackage[breaklinks=true]{hyperref}
\usepackage{graphicx}
\usepackage[english]{babel}
\usepackage{amsmath}
\usepackage{amssymb}
\usepackage{times}

\begin{document}

\newcommand{\D}{\mathrm{D}}
\newcommand{\p}{\partial}
\newcommand{\Tr}{\mathrm{Tr}}
\renewcommand{\d}{\mathrm{d}}
\newcommand{\Ek}{E_\mathbf{k}}
\newcommand{\xik}{\xi_\mathbf{k}}
\newcommand{\sumk}{\sum_\mathbf{k}}

\title{Pairing phenomena and superfluidity of atomic Fermi gases in a two-dimensional optical lattice: Unusual effects of lattice-continuum mixing}

\author{Lin Sun}
\affiliation{Department of Physics and Zhejiang Institute of Modern Physics, Zhejiang University, Hangzhou, Zhejiang 310027, China}

\author{Jibiao Wang}
\affiliation{School of Materials and New Energy, South China Normal University, Shanwei, Guangdong 516625, China}

\author{Xiang Chu}
\affiliation{Department of Physics and Zhejiang Institute of Modern Physics, Zhejiang University, Hangzhou, Zhejiang 310027, China}

\author{Qijin Chen}
\email[Corresponding author: ]{qchen@uchicago.edu}
 \affiliation{Hefei National Laboratory for Physical Sciences at Microscale and School of Physics, University of Science and Technology of China,  Hefei, Anhui 230026, China}
\affiliation{Shanghai Branch, CAS Center for Excellence and Synergetic Innovation
  Center in Quantum Information and Quantum Physics, University of
  Science and Technology of China, Shanghai 201315, China}
\affiliation{Shanghai Research Center for Quantum Sciences, Shanghai 201315, China}
\affiliation{Department of Physics and Zhejiang Institute of Modern Physics, Zhejiang University, Hangzhou, Zhejiang 310027, China}

\date{\today}

\begin{abstract}
  We study the superfluid behavior of ultracold atomic Fermi gases
  with a short range attractive interaction in a two-dimensional
  optical lattice (2DOL) using a pairing fluctuation theory, within
  the context of BCS-BEC crossover.  We find that the mixing of
  lattice and continuum dimensions leads to exotic phenomena. For
  relatively large lattice constant $d$ and small hopping integral
  $t$, the superfluid transition temperature $T_c$ exhibits a
  remarkable reentrant behavior as a function of the interaction
  strength, and leads to a pair density wave ground state, where $T_c$
  vanishes, for a range of intermediate coupling strength. In the
  unitary and BCS regimes, the nature of the in-plane and overall
  pairing changes from particle-like to hole-like, with an unexpected
  nonmonotonic dependence of the chemical potential on the pairing
  strength.  The BEC asymptotic behaviors exhibit distinct power law
  dependencies on the interaction strength compared to cases of pure
  3D lattice, 3D continuum, and 1DOL. These predictions can be tested in future experiments. \hspace*{1.1cm} {\sffamily Ann Phys (Berlin) {\bf 2022}, 2100511}
\end{abstract}

\maketitle

\section{Introduction}

Ultracold Fermi gases have become a powerful platform for
investigating condensed matter physics over the past decade, due to
their multiple experimentally tunable control parameters
\cite{chen2005PR,bloch2008RMP}.  They can exhibit a perfect crossover
from a BCS type of superfluidity to Bose-Einstein condensation (BEC)
of atom pairs, as the pairing strength is tuned from weak to strong
via a Feshbach resonance \cite{chin2010RMP}.  In particular, they can
be put in an optical lattice with a tunable geometry so that rich
physics may emerge \cite{bloch2005NP,kohl2005PRL,schneider2008S}.  In
a three dimensional (3D) lattice, it is known that the superfluid
transition temperature $T_c$ scales as $-t^2/U$ in the BEC regime, due
to ``virtual ionization'' in the pair hopping process
\cite{NSR,chen1999PRB}, where $U<0$ is the onsite attractive
interaction, and $t$ is the hopping integral.  This is to be
contrasted with the 3D continuum case, in which $T_c$ approaches
asymptotically $0.218T_F$ in the BEC regime, with $T_F$ being the
Fermi temperature. Given the sharp contrast between the 3D continuum
and the 3D lattice cases, usual behaviors may be expected when the
spatial dimensions contain both lattice and continuum. Recently, as
such a lattice-continuum mixed system, the superfluid behavior of
Fermi gases in a 1D optical lattice (1DOL), in the context of BCS-BEC
crossover, has been studied. Interesting effects of lattice-continuum
mixing have indeed been found. For example, in the population balanced case, the
behavior of $T_c$ exhibits lattice effect; $T_c$ decreases with
increasing coupling strength in the BEC regime, with a scaling of
$T_c = \pi an/2m$, where $a$ is the $s$-wave scattering length, $n$
the fermion density and $m$ the fermion mass \cite{wang2020PRA}. In
contrast, in the presence of population imbalance \cite{wang2020PRA2},
$T_c$ often approaches a constant BEC asymptote, mimicking its
counterpart in the 3D continuum.

Two dimensional optical lattices (2DOL) have also been studied
theoretically \cite{parish2007PRL,roscher2014PRA,zhao2008PRA}. They
corresponds to a 2D array (2D lattice in the $xy$ plane) of 1D tubes
(1D continuum alone the $z$ direction) in experiment, with lattice
constant $d$ and hopping integral $t$, subject to a short-range
pairing interaction $U (< 0)$.  2DOL of $^6$Li in the (quasi-)1D limit
has been realized experimentally
\cite{liao2010N,pagano2014NP,revelle2016PRL}. None of these
theoretical and experimental studies paid attention to or was aware of
possible lattice-continuum mixing effects in a more general parameter
range. Most of them have focused on the 1D or quasi-1D physics,
including the Fulde-Ferrell-Larkin-Ovchinnikov states (FFLO) state
\cite{FF,*LO} in the presence of a population imbalance. Dimensional
crossover was also studied \cite{revelle2016PRL} in a population
imbalanced Fermi gas of $^6$Li atoms, with a focus on how phase
separation evolves with $t$ and the interactions. Apparently, the
lattice-continuum effects may not show up for any arbitrary
choices of $d$ and $t$ as well as interaction strength.

In this paper, we investigate the superfluid behavior of a balanced
two-component Fermi gas in a 2DOL. We find a much stronger effect of
lattice-continuum mixing than in the 1DOL case, besides the wide
spread pseudogap phenomena \cite{chen2005PR}. In particular, two
lattice dimensions provide a tighter confinement in the momentum
space. Via increasing lattice constant $d$ and decreasing hopping
integral $t$, such that the Fermi surface becomes open (in contrast to a 3D closed ellipsoid), the effective number density in the lattice dimensions
can go above half filling, and may turn the in-plane and overall
pairing from particle-like into hole-like in the BCS and unitary
regimes, with chemical potential $\mu$ increasing with increasing
pairing strength, while it remains particle-like in the BEC regime.
While the decrease of $T_c$ in the BEC regime manifests strong lattice
effects, we find that for large $d$ and small $t$, $T_c$ exhibits a
re-entrant behavior in the regime of intermediate pairing strength; a
pair density wave (PDW) emerges as the ground state where $T_c$
vanishes. In the BEC regime, the momentum space becomes quasi-1D, so
that $T_c$ approaches $ \frac{\sqrt{6}ntd^4}{m}|U|^{-1}$
asymptotically, while $\mu \sim -U^2$ is dominated by a quadratic
dependence on $U$, with the pairing gap
$\Delta \sim |\mu|^{3/4} \sim |U|^{3/2}$. These behaviors are distinct
from their counterpart in 3D lattice, 3D continuum or 1DOL.
  
It is known that the superfluid phase in Fermi gases in 3D optical
lattices has yet not been accessible experimentally, due to the
difficulty of deep cooling. On the contrary, superfluidity in 1D or 2D
optical lattices have been realized long ago, e.g., in the early phase separation observation  \cite{Rice1}.

\section{Theoretical Formalism}
\subsection{Pairing ﬂuctuation theory}

To focus on the effects of lattice-continuum mixing, we shall focus on
the 2DOL without the complication of population imbalance.  Under the
one-band tight-binding approximation for the lattice dimensions in the
$xy$ plane, the dispersion of noninteracting atoms is given by
$\xik=\epsilon_\mathbf{k}-\mu \equiv{k_z}^2/2m + 2t[2-\cos (k_xd)-\cos
(k_yd)]-\mu$.
Following our recent studies
\cite{Zhang17SR,UltraHighTc,wang2020PRA,wang2020PRA2}, we take $t$ to
be small in our calculation, under the constraint $2mtd^2 < 1$, so
that it is physically accessible. The critical coupling strength,
which corresponds to a diverging $s$-wave scattering length $a$ in the
unitary limit in the presence of the lattice, is given by
$U_c=-1/\sum_{\textbf{k}}1/2\epsilon_{\textbf{k}}=-0.16072\sqrt{2m}/\sqrt{t}d^2$. Note
that a big difference between 2DOL and 1DOL as well as 3D continuum is
that $U_c=0$ in the contact potential limit for the latter two
cases. As a consequence, $1/k_Fa$ as defined via the Lippman-Schwinger
equation is upper bounded in the present case. Thus 2DOL is closer to
the 3D lattice case in this sense. In comparison with the pure 3D
lattice case, however, the momentum distribution will spread to large $k_z$ (in the BEC case or for small $t$  and large $d$),
leading to a quasi-1D momentum space, with the kinetic energy
distributed mainly along the $k_z$ direction. For  3D lattice,
which has a finite momentum space defined by the first Brillouin zone (BZ),
there is a finite filling factor for a given fermion density. However,
the overall filling factor is zero for 2DOL. Nevertheless, as we show
below, one can tune $t$ and $d$ such that the effective filling factor
within the two lattice dimensions  varies between 0 and 1.

With the modified atomic dispersion $\xik$ for 2DOL, we follow the pairing
fluctuation theory previously developed for the pseudogap physics in
the cuprates \cite{chen1998PRL}, which has been extended to address
the BCS-BEC crossover in ultracold atomic Fermi gases
\cite{chen2005PR}.  This theory goes beyond the BCS mean-field
approximation by self-consistently including finite momentum pairing
correlations in the single particle self energy.  Thus the self energy
$\Sigma(K) = \Sigma_{sc}(K) + \Sigma_{pg}(K)$ contains two parts,
where the Cooper pair condensate contribution
$\Sigma_{sc}(K) = -\Delta_{sc}^2G_0(-K)$ vanishes above $T_c$, and
finite momentum pair contribution
$\Sigma_{pg}(K) = \sum_Q t_{pg}(Q)G_0(Q-K)$ exists both above and
below $T_c$.
Here $G_0$ and $G$ are the non-interacting and full Green's functions, respectively, $\Delta_{sc}$
the order parameter and $t_{pg}(Q)= U/[1+U\chi(Q)]$ the pairing $T$ matrix, with pair susceptibility $\chi(Q)=\sum_K G_0(Q-K)G(K)$. We take
$\hbar=k_B=1$ and use four momenta
$K\equiv (\omega_n, \mathbf{k})$, $Q\equiv (\Omega_l, \mathbf{q})$,
$\sum_Q\equiv T\sum_l\sum_\mathbf{q}$, etc, where $\omega_n$
($\Omega_l$) is the odd (even) Matsubara frequency, following the
notations of Ref.~\cite{chen1998PRL} \footnote{We also drop the volume $V$, which can be set to unity and will be canceled out in the end.}. 

We now recapitulate the main result of this theory. In the superfluid
phase, one can define the pseudogap via
$\Delta_{pg}^2 = -\sum_Q t_{pg}(Q)$, so that the total gap $\Delta$ is
given by $\Delta^2 = \Delta_{sc}^2 + \Delta_{pg}^2$, which leads to
the self energy $\Sigma(K) \approx -\Delta^2 G_0(-K)$, and the BCS form of the full
Green's function 
$
 G(K)=\frac{u_{\textbf{k}}^{2}}{i\omega_{n}-E_{\textbf{k}}}+\frac{v_{\textbf{k}}^{2}}{i\omega_{n}+E_{\textbf{k}}}\,,
$
where $u_{\textbf{k}}^{2}=(1+\xi_{\textbf{k}}/E_{\textbf{k}})/2$,
$v_{\textbf{k}}^{2}=(1-\xi_{\textbf{k}}/E_{\textbf{k}})/2$, 
and $E_{\textbf{k}}=\sqrt{\xi_{\textbf{k}}^{2}+\Delta^{2}}$. 
Then we have the number equation,
\begin{equation}
  n=\sum_{\textbf{k}}\Big[\Big(1-\frac{\xi_{\textbf{k}}}{E_{\textbf{k}}}\Big)
 +2\frac{\xi_{\textbf{k}}}{E_{\textbf{k}}}f(E_{\textbf{k}})\Big]\,, 
\label{eq:neq}\\
\end{equation}
where $f(x)$ is the Fermi distribution function.
At $T \le T_c$, the Thouless criteria  for pairing instability, $U^{-1}+\chi(0)=0$, leads to the gap equation, 
%
\begin{equation}
  0=\frac{1}{U}+\sum_{\textbf{k}}\Big[\frac{1-2f(E_{\textbf{k}})}{2E_{\textbf{k}}}\Big]\,.
  \label{eq:gap}
\end{equation}

To evaluate the pseudogap, one can 
Taylor expand $t_{pg}^{-1}(Q)$ on the real frequency axis, after analytical continuation,
$t_{pg}^{-1}(\Omega,\textbf{q})\approx a_1\Omega^{2}+a_{0}(\Omega-\Omega_{\textbf{q}}+\mu_{p})$, with
$\Omega_{\textbf{q}}=B{q_z}^{2} + 2t_{B}[2-\cos(q_{x}d)-\cos(q_{y}d)]$, and $\mu_p=0$ at $T\le T_c$. 
Here $B=1/2M$, with $M$ being the effective pair mass in the $z$ direction, 
and $t_{B}$ the effective pair hopping integral in the $xy$ plane.
We stress that the $a_1$ term serves as a small quantitative correction in the BEC regime, where  $a_1T_c \ll a_0$. 
Then this $T$-matrix expansion leads to the pseudogap equation
\begin{equation}
  |a_0|\Delta_{pg}^{2}=\sum_{\textbf{q}}\frac{b(\tilde{\Omega}_{\textbf{q}})}{\sqrt{1+4\dfrac{a_{1}}{a_{0}}\Omega_{\textbf{q}}}}\,,
  \label{eq:PG}
\end{equation}
where $b(x)$ is the Bose distribution function and
$\tilde{\Omega}_{\textbf{q}}=[\sqrt{a_0^{2}+4a_1a_0\Omega_{\textbf{q}}}-a_0]/2a_1$
is the pair dispersion, which reduces to
$\tilde{\Omega}_{\textbf{q}}=\Omega_{\textbf{q}}$ when
$a_1T_c/a_0 \ll 1$, e.g., in the BEC regime.  In this case, one can
see that the pair density is roughly given by $n_p = a_0 \Delta^2$

Equations (\ref{eq:neq})-(\ref{eq:PG}) form a closed set of self-consistent equations, which can be used to solve for
($\mu$, $\Delta_{pg}$, $T_c$) with $\Delta_{sc}=0$.
For our numerics, we define Fermi momentum $k_{F}=(3\pi^{2}n)^{1/3}$ 
and Fermi energy $E_{F}\equiv k_{B}T_{F}=\hbar^{2}k_{F}^{2}/2m$, as for the 3D continuum with the same density $n$ and atomic mass $m$.
  

\subsection{Asymptotic behaviors in the BEC limit}
The asymptotic behavior at $T_c$ in the BEC limit can be solved
analytically, with $\mu \rightarrow -\infty$, so that
$f(\Ek) = f(\xik) = 0$.
From the gap equation, we obtain
\begin{equation}
  \mu = -\frac{m}{8d^4}U^2+\frac{3n}{4}|U|\,.
  \label{eq:muBEC}
\end{equation}
The leading order quadratic dependence   of $\mu$ on $U$ is related to the
quasi-one dimensionality caused by the restricted momentum space for
($k_x$,$k_y$) within the first BZ. This should be contrasted to the 3D lattice case, for which $\mu \sim U$ in the BEC limit \footnote{For 1DOL and 3D continuum, $U$ is renormalized down to zero in the contact limit so that an appropriate parameter is scattering length $a$. We have $\mu \sim - e^{d/a}$ and $\mu\sim -1/a^2$ for the two cases, respectively.}.
From the number equation, we get
\begin{equation}
  \Delta = \left(\frac{8d^2n}{\sqrt{2m}}\right)^{1/2}|\mu|^{3/4}\,.
  \label{eq:gapBEC}
\end{equation}
This $\Delta\sim |\mu|^{3/4}\sim |U|^{3/2}$ scaling behavior
should be contrasted to  $\Delta \sim \mu \sim |U|$ for 3D lattice, $\Delta\sim |\mu|^{1/2}$ for 1DOL  (quasi-2D system) \cite{wang2020PRA}, and
 $\Delta\sim |\mu|^{1/4}$  for 3D continuum \cite{chen2005PR}.

For the effective pair mass in the continuum dimension and the effective pair hopping integral in the lattice dimensions, one can easily obtain
\begin{equation}
   B = \frac{1}{4m}, \quad \mbox{and}\quad
   t_B = \frac{3t^2}{4|\mu|},
   \label{eq:tBBEC}
\end{equation}
which yields
\begin{equation}
  n \approx \frac{\sqrt{T_c}}{\sqrt{\pi B}d^2}Li_{1/2}(e^{-4t_B/T_c})
\end{equation}
in the BEC regime via the pseudogap equation (\ref{eq:PG}), where $Li_{s}(z)=\sum_{n=1}^{\infty}(z^n/n^{s})$ 
is the polylogarithm function.
Note here the $a_1$ term is negligible, so that $n_p = a_0\Delta^2 = n/2$.
Thus we obtain the BEC  asymptotic solution 
\begin{equation}
  T_c \approx 2nd^2\sqrt{Bt_B} \approx \frac{\sqrt{6}ntd^4}{m}|U|^{-1}
  \label{eq:TcBEC}
\end{equation}
using the lowest order expansion, $Li_{1/2}^{-2}(1-z) \approx z/\pi$ for $z\rightarrow 0$.
This indicates that $T_c$  scales inversely proportional to $U$ in the deep BEC regime, similar to the 3D lattice case.

\section{Results and Discussions}

\begin{figure}
  \includegraphics[clip,width=3.4in]{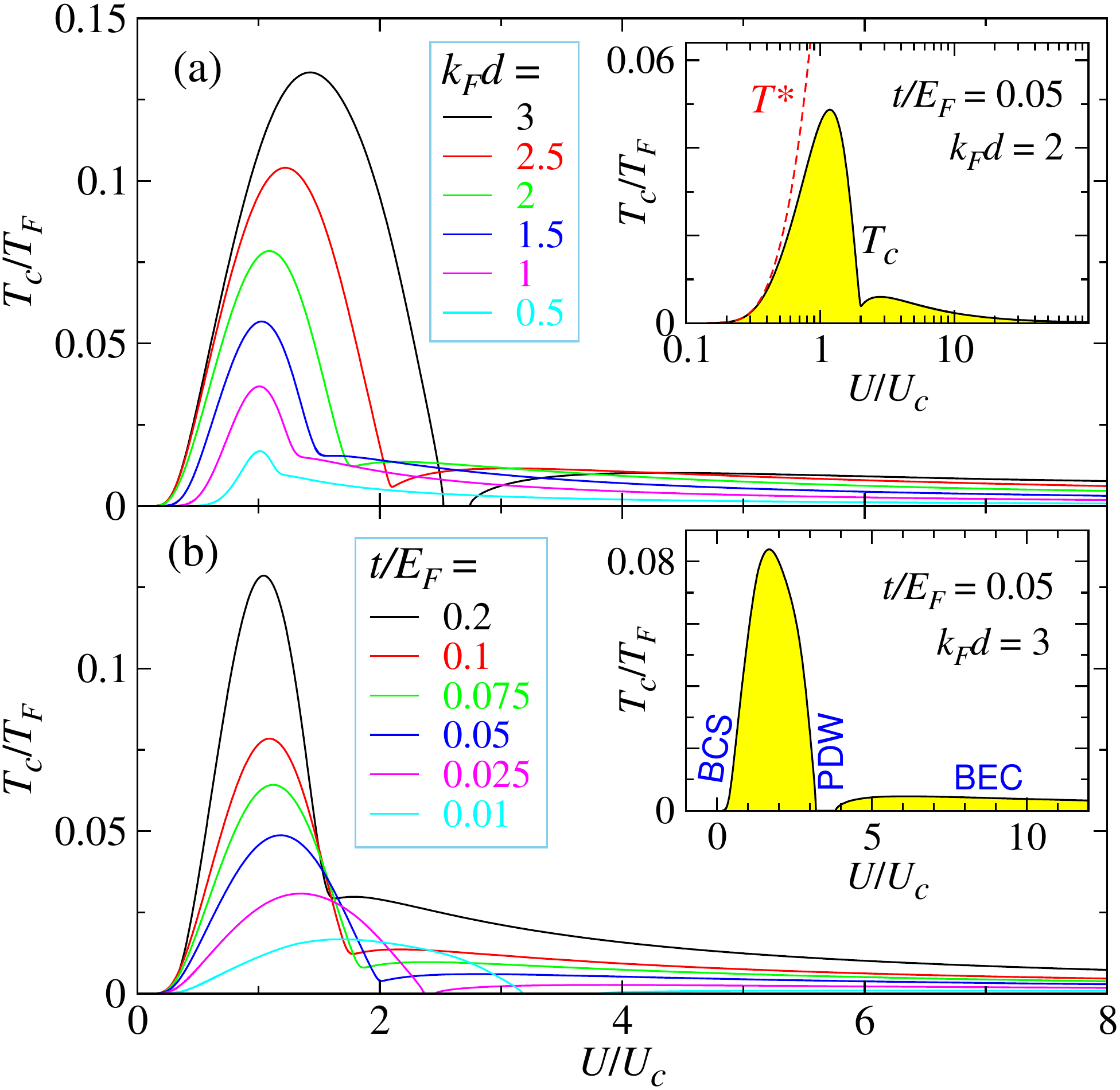}
  \caption{Phase diagram and effect of varying $d$ and $t$: Behavior of $T_c$ versus $U/U_c$ with (a) fixed $t/E_F=0.1$ for varying $k_Fd$, 
    and (b) fixed $k_Fd=2$ for varying $t/E_F$, as labeled. The insets of (a) and (b) show typical phase diagrams without and with a PDW phase, respectively. }
  \label{fig:Tc1}
\end{figure}

Shown in the inset of Fig.~\ref{fig:Tc1}(a) is a typical phase diagram of Fermi gases in 2DOL,  with $t/E_F=0.05$ and $k_Fd=2$, which satisfies  $2mtd^2=0.2 < 1$. A semi-log scale is used to show the full interaction range.
The black solid and red dashed curves are $T_c$ and the pair formation
temperature $T^*$ (which is approximated by the mean-field solution
for $T_c$), respectively, so that the yellow shaded area indicates the
superfluid phase.  A pseudogap state exists between $T^*$ and
$T_c$.  As usual, $T_c$ follows the BCS solution in the weak-coupling regime.  Then a pseudogap emerges as $|U|$ increases, which suppresses $T_c$ relative to $T^*$. After reaching a 
maximum around unitarity at $U=U_c$, $T_c$ starts to decrease due to a shrinking Fermi surface. A minimum is reached when $\mu=0$, where the Fermi surface shrinks to zero and essentially all fermions have paired up. Beyond this point, as the pairing becomes stronger, the system enters the bosonic regime,
 with $\mu < 0$, and $T_c$ first rises due to shrinking pair size and hence decreasing pair mass, and then decreases 
 towards the BEC asymptote, as governed by the lattice effect \cite{NSR,chen1999PRB}.

 We note that a constant Hartree energy, $\Sigma_\text{HF} = Un/2$, has been absorbed in our chemical potential $\mu$ \footnote{See Eq.~(2.4) of Ref.~\cite{ChenPhD}}. The actual chemical potential  is given by $\mu'= \mu + Un/2$. Nevertheless, we emphasize that the vanishing of Fermi surface is defined by $\mu=0$ rather than $\mu'=0$. \footnote{In principle, $\mu$, when positive, can be measured experimentally by detecting  $k_\mu = \sqrt{2m\mu}$ (in the $z$ direction) on the Fermi surface directly via measurement of the fermion momentum distribution, e.g., using the time-of-flight imaging technique.}

To demonstrate the effect of varying $d$ and $t$, we plot in
Fig.~\ref{fig:Tc1} $T_c$ versus $U/U_c$ with (a) fixed $t/E_F=0.1$ but
for varying $k_Fd$ from 0.5 through 3, and (b) fixed $k_Fd=2$ but for
varying $t/E_F$ from 0.01 through 0.2.
For fixed $t$ in panel (a), the maximum $T_c$,
${T_c}^\text{max}$, increases with $d$.  At the same time, the entire
$T_c$ curve also expands horizontally, as $d$ increases.  Since the
lattice momentum ($k_x,k_y$) is restricted to the first BZ, fermions almost fully occupy the $xy$-plane states
when  $t$ is small \cite{wang2020PRA}.  As $d$ increases, the BZ shrinks, and more fermions have to occupy higher $k_z$ state, so that $\mu$ is pushed up, leading to higher $T_c$. 
Similarly in panel (b),  as $t$ decreases with fixed $d$, ${T_c}^\text{max}$ decreases and moves toward the BEC side of unitarity, 
as if the entire $T_c$ curve is expanding horizontally toward the BEC regime.
For fixed $d$, more fermions will occupy the first BZ in the $xy$ plane, as $t$ decreases. This will remove the high $k_z$ fermions, and decrease $\mu$. Meanwhile, a small  $t$ causes difficulty for pair hopping and thus suppresses $T_c$, making the system more 1D.

Remarkably, we can see that $T_c$ exhibits a reentrant behavior for large $d$ and small $t$. This is illustrated more clearly as an example in the inset of Fig.~\ref{fig:Tc1}(b) for $k_Fd=3$ and $t/E_F=0.05$.
For a range of intermediate coupling strength, $T_c$ shuts off completely before it recovers at stronger couplings in the BEC regime, 
and PDW states emerge where $T_c$ vanishes, 
which is also found in dipolar Fermi gases within our pairing fluctuation theory \cite{che2016PRA}. 
\begin{figure}
  \includegraphics[clip,width=3.4in]{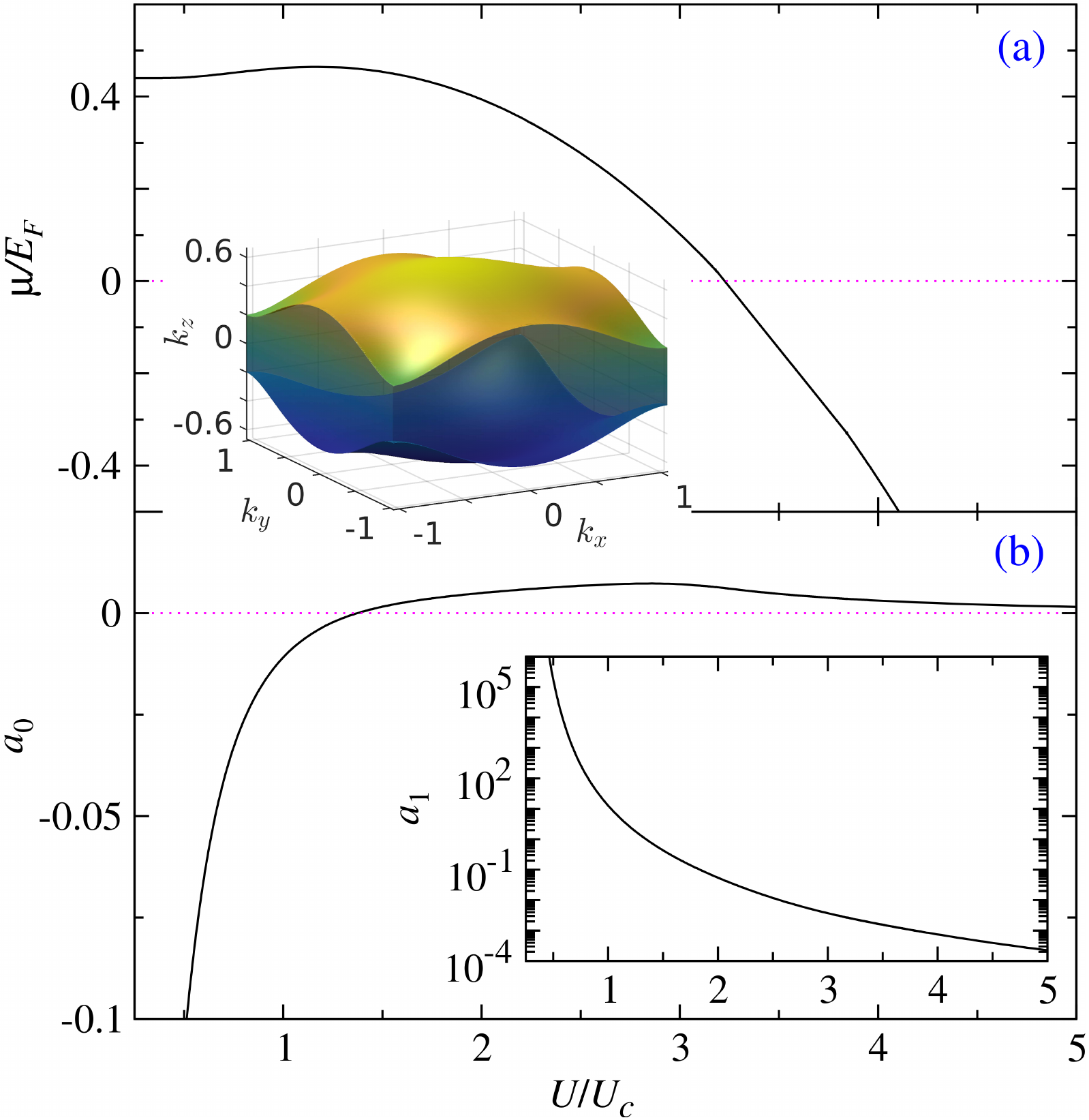}
  \caption{Evolution of (a) $\mu$ and (b) the coefficients $a_0$ and $a_1$ (inset) at $T_c$ as a function of  $U/U_c$  for $t/E_F=0.05$ and $k_Fd=3$. $T=0$ is used in the PDW state. Also shown is the 3D plot of the corresponding Fermi surface. 
}
  \label{fig:a0}
\end{figure}

\begin{figure}
  \includegraphics[clip,width=3.4in]{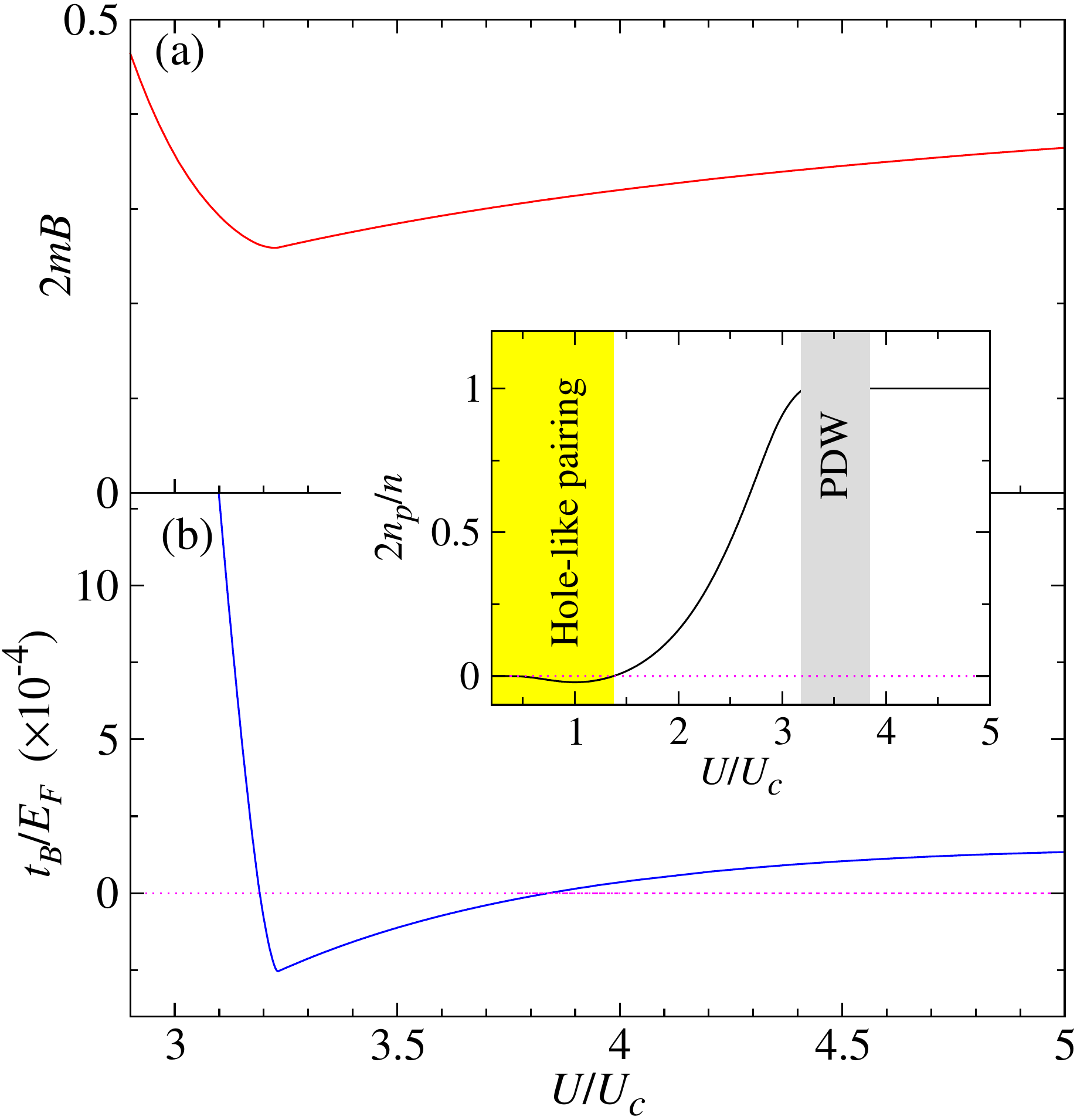}
  \caption{Behavior of (a) $B$ and (b) $t_B$ as well as $2n_p/n$ (inset) as a function of  $U/U_c$, calculated with the same parameters as in Fig.~\ref{fig:a0}. The yellow and grey shaded areas indicate hole-like pairing (with $a_0<0$) and PWD state, respectively. }
  \label{fig:Tc2}
\end{figure}

Shown in Fig.~\ref{fig:a0} are the behaviors of (a) $\mu$ and (b) the
coefficients $a_0$ and $a_1$ (inset) as a function of $U/U_c$ for
$t/E_F=0.05$ and $k_Fd=3$.  It is remarkable that $\mu$ increases
first as the pairing becomes stronger, which is distinct from the 3D
continuum and 3D lattice cases below half filling. Even more
remarkable is that $a_0 < 0$ before $\mu$ reaches its maximum near
$U/U_c = 1.38$, even though the monotonic behavior of $a_1$ is rather
conventional. For larger $U/U_c$, $a_0$ becomes positive, and then
starts to decrease with $U/U_c$ in the BEC regime, where $\mu$ becomes
negative.  The \emph{unusual} behaviors of $\mu$ and $a_0$ can be
explained by the Fermi surface, as shown in the inset. It has open
ends, in contrast to a closed ellipsoid for cases of large $t$ and
small $d$. As is well known, in the 3D lattice case above half
filling, the Fermi surface becomes hole-like, for which $a_0$ is
negative and the chemical potential $\mu$ increases as pairing
strength increases. This can be easily understood via a particle-hole
transformation. Therefore, the sign of $a_0$ determines whether the
fermion pairs are particle-like or hole-like. The negative sign has to
do with the negative quasiparticle velocity on the hole-like Fermi
surface. For the present 2DOL case shown in Fig.~\ref{fig:a0}, the
effective filling factor in the first BZ of the $xy$ plane is above
1/2, so that the contributions from lattice dimensions dominate the
sign of $a_0$ in the unitary and weak-coupling regimes, even though
the pairing in the $z$-direction always remains particle-like. In the
BEC regime, the gap becomes large so that the contributions of the
lattice dimensions spread across the entire BZ and become less
dominant. As a result, the continuum dimension will dominate the sign
of $a_0$ in the BEC regime.  For hole-like pairing (as shown in the
yellow region in the inset of Fig.~\ref{fig:Tc2}), we have $a_0<0$ and
thus $n_p <0$, as required by the fact $\mu$ is larger than its
noninteracting counterpart, via the general relation
$n_p = n/2 - \sumk f(\xik)$.

\begin{figure}
  \includegraphics[clip,width=3.3in]{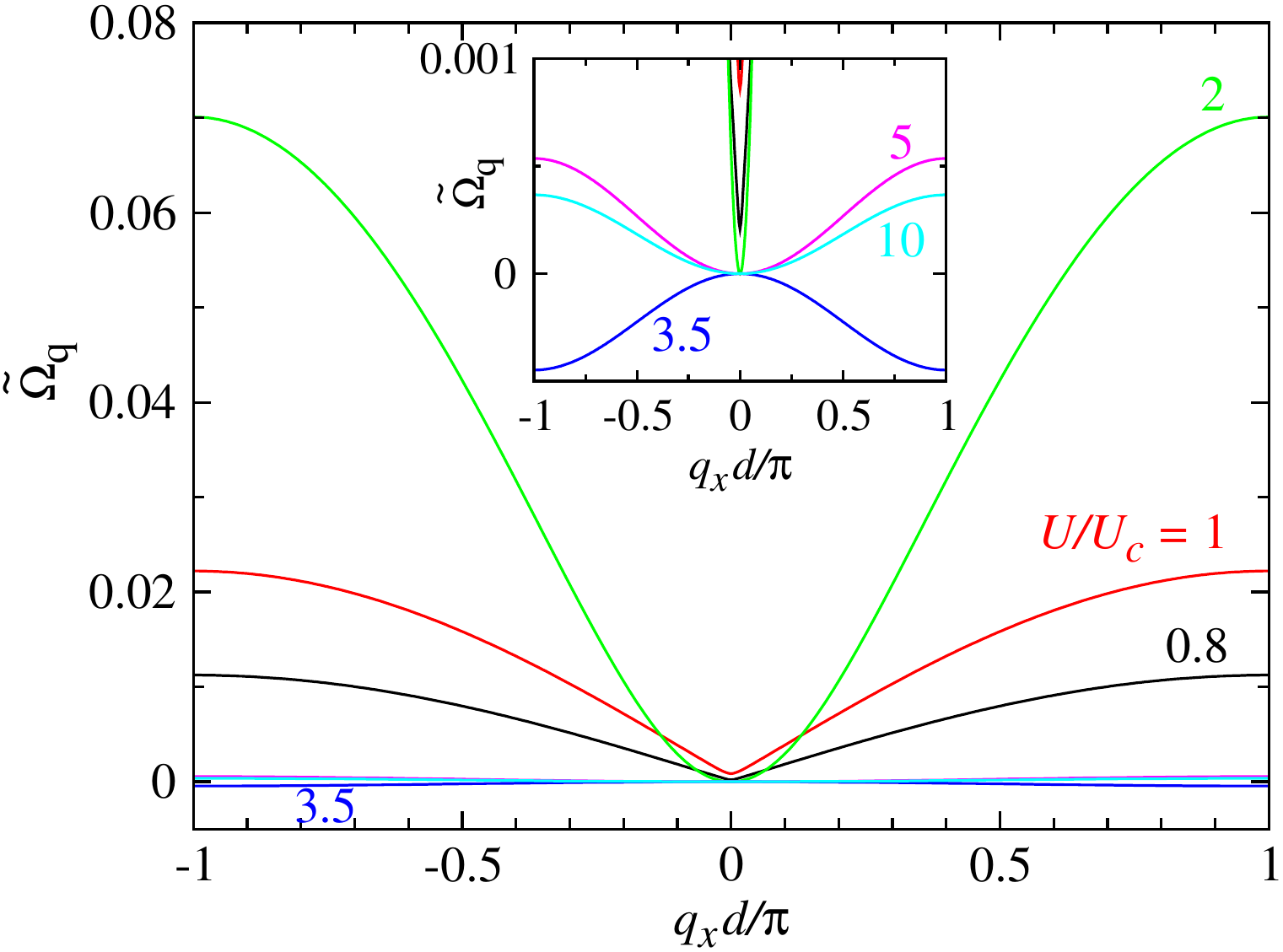}
  \caption{Pair dispersion $\tilde{\Omega}_{\textbf{q}=(q_x,0,0)}$  along the $q_x$ direction for different $U/U_c$, as labeled, calculated with the same parameters as in Fig.~\ref{fig:a0}.  For $U/U_c=3.5$, the $\Gamma$ point, $(q_x,q_y,q_z)=0$, becomes a local maximum of $\tilde{\Omega}_{\textbf{q}=(q_x,0,0)}$  in the $q_x$ and $q_y$ directions, signaling an instability of the conventional pair condensation at $q=0$, which hence leads to a PDW state.}
  \label{fig:Omegaq}
\end{figure}

In Fig.~\ref{fig:Tc2}, we show the corresponding behaviors of (a) the inverse pair mass $B$ and (b) the pairing hopping integral $t_B$ as well as pair fraction $2n_p/n$ (inset) as a function of  $U/U_c$ for $t/E_F=0.05$ and $k_Fd=3$, calculated at $T_c$. We set $T=0$ for the  PDW states. The behavior of $B$ is rather conventional, with a minimum around $\mu=0$. In contrast, $t_B$ is negative in the PDW state, leading to a vanishing $T_c$, at a intermediate coupling strength, $3.19\leq U/U_c\leq 3.85$. Upon the sign change of $t_B$ from positive to negative, the minimum of $\tilde{\Omega}_{\textbf{q}}$  will literally move from $\mathbf{q}=0$ to $\mathbf{q} = (\pi/d,\pi/d,0)$, within the lowest order tight-binding approximation for the in-plane pair dispersion, $\tilde{\Omega}_{\textbf{q}=(q_x,q_y,0)}$, as shown by the $U/U_c=3.5$ case in  Fig.~\ref{fig:Omegaq} (along the $q_x$ direction).
This corresponds to a first order phase transition from a homogeneous
superfluid to a PDW ground
state. 
\footnote{When higher order wavevectors are included in the pair dispersion, the minimum may be shifted to  a different location.}
The physical origin of the PDW states has to do with the strong inter-pair repulsive interaction for relatively large pair size and moderately strong pairing interaction, which leads to Wigner crystallization in the $xy$ plane. 
It remains unclear and will be left for future studies  whether the PDW phase can sustain superfluidity and thus becomes a supersolid.
 Note that $\tilde{\Omega}_{\textbf{q}}$ becomes gapped when $a_0<0$, which reduces the pseudogap.

\begin{figure}
  \includegraphics[clip,width=3.2in]{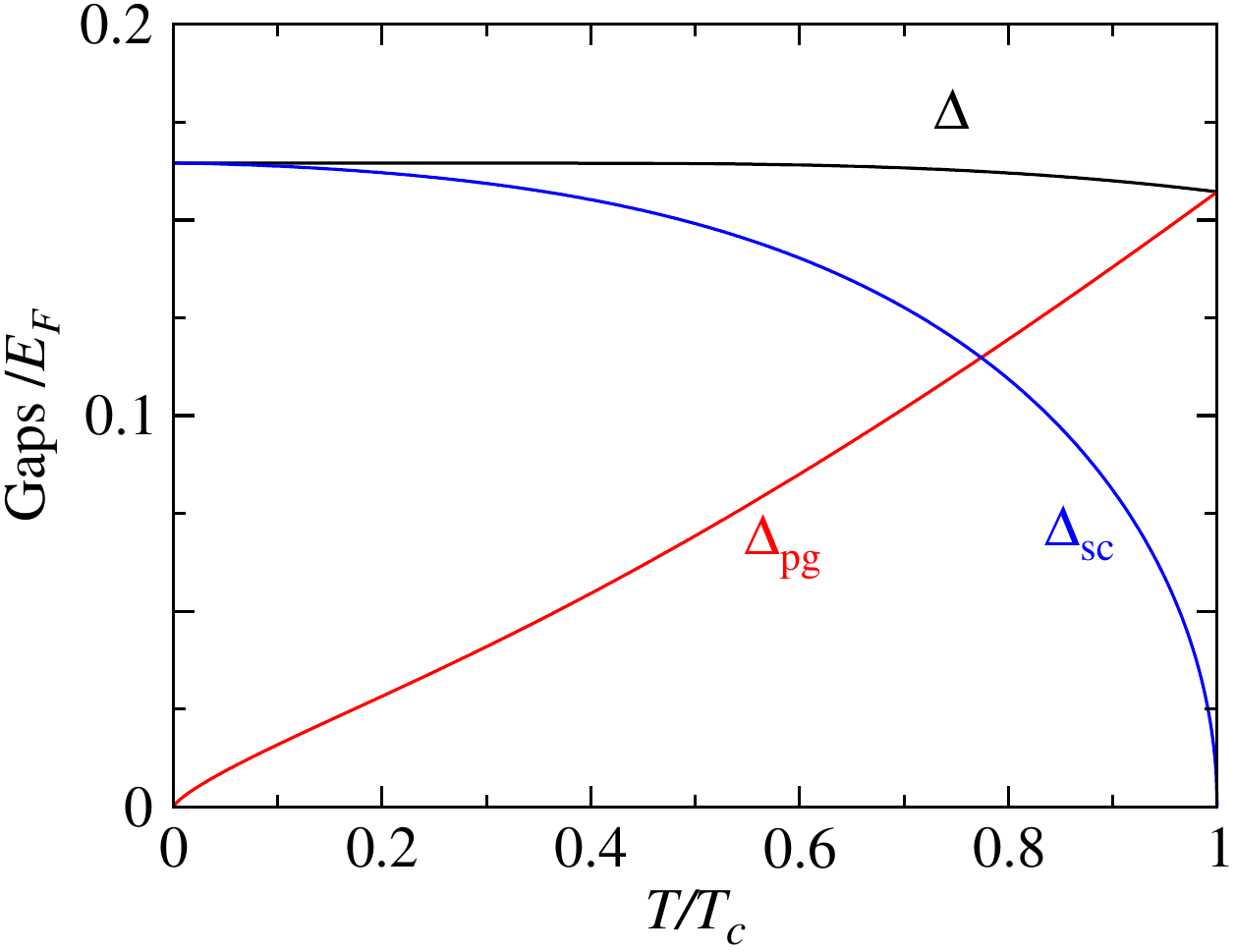}
  \caption{Temperature dependence of the gaps and the order parameter at and below $T_c$ at unitarity, $U/U_c=1$, for $k_Fd=2$ and $t/E_F=0.05$. The order parameter $\Delta_{sc}$ decreases with $T$ and necessarily closes at $T_c$. At very low $T$, $\Delta_{pg}^2\propto n_\text{pair} \propto T^{3/2}$ so that $\Delta_{pg}\propto T^{3/4}$, which increases with $T$. Note that there is already a rather large (pseudo)gap at $T_c$ in the unitary regime. }.
  \label{fig:Gaps}
\end{figure}

\begin{figure}
  \includegraphics[clip,width=3.4in]{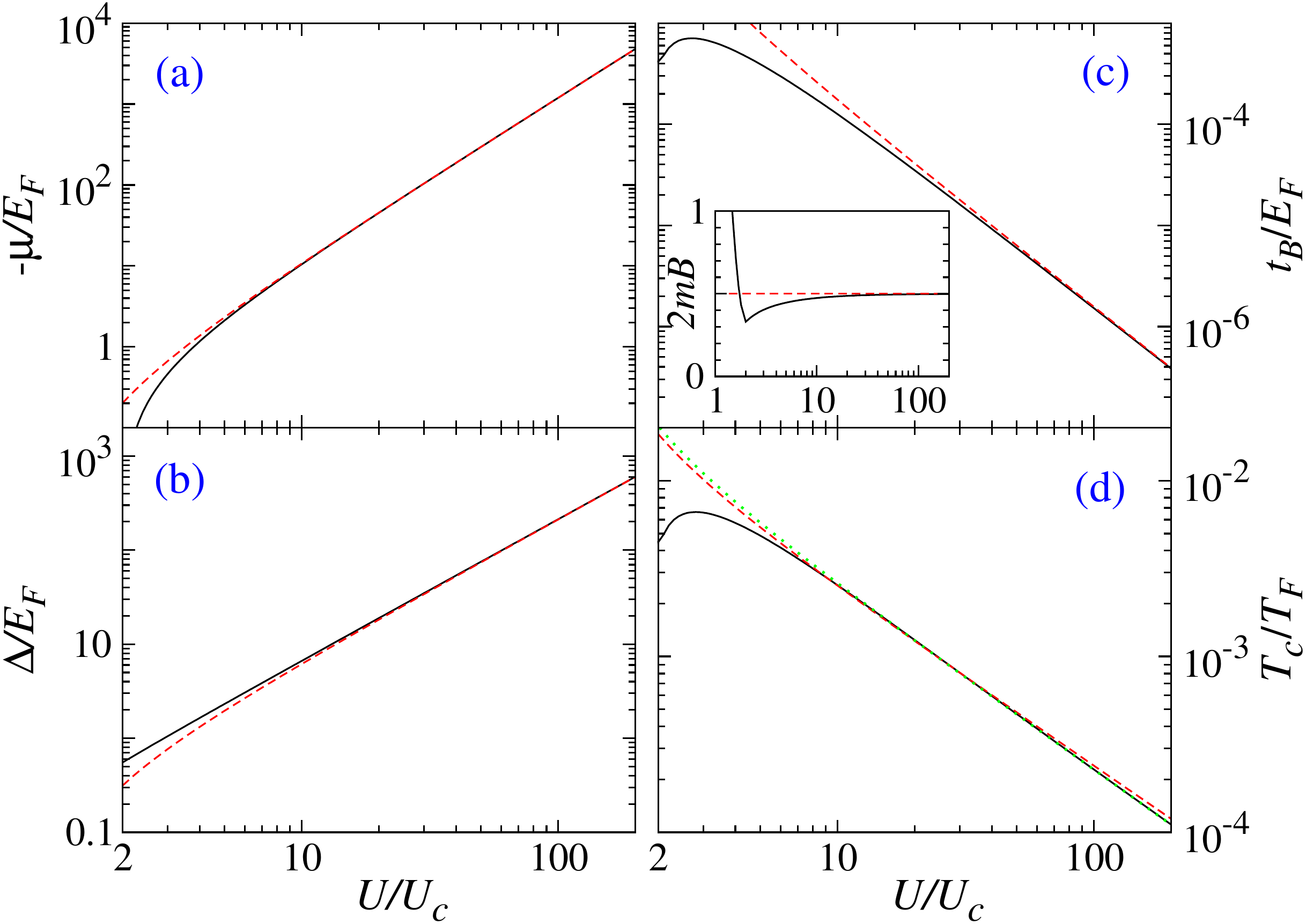}
  \caption{Comparison between full numerical solutions (black solid)
    and BEC asymptotic behaviors (red dashed lines) of (a) $-\mu$, (b)
    $\Delta$,(c) $t_B$ and $B$ (inset), and (d) $T_c$ as a function of
    $U/U_c$ on a log-log scale, for $t/E_F=0.05$ and $k_Fd=2$. The green dotted line contains higher order contributions in the expansion of $Li_{1/2}^{-2}(1-z)$ }.
  \label{fig:BEClimit}
\end{figure}

Next, we show in Fig.~\ref{fig:Gaps} the typical behaviors of the order parameter $\Delta_\text{sc}$ and the gaps $\Delta$ and $\Delta_\text{pg}$ as a function of temperature at unitarity. They are qualitatively similar to their counterpart in the 3D continuum and 3D lattice cases. As $T$ increases from zero, the order parameter decreases and vanishes at $T_c$, while the pseudgap increases from zero and reaches its maximum at $T_c$. The increase of  $\Delta_\text{pg}$ reflects the nature that it results from finite momentum pairing, whose population depends on thermal activation. The combined total gap $\Delta$ follows the BCS mean-field-like behavior, and would continue to exist up to $T^*$ far above $T_c$. As a function of interaction strength, as $T_c$ decreases rapidly with decreasing $U/U_c$ towards the BCS regime, all gaps decrease and the ratio $\Delta_\text{pg}(T_c)/\Delta(0)$ decreases and becomes essentially zero in the extreme BCS limit. In contrast, with increasing  $U/U_c$ towards the BEC regime, this ratio approaches unity, while all gaps increase. In this case, $\Delta$ becomes essentially temperature  independent, and system can be regarded as a composite Bose gase.

Finally, we show in Fig.~\ref{fig:BEClimit} the BEC asymptotic behavior (red
dashed lines) of (a) $-\mu$, (b) $\Delta$, (c) $t_B$ and $B$ (inset)
and (d) $T_c$ as a function of $U/U_c$ on a log-log scale for
$t/E_F=0.05$ and $k_Fd=2$, obtained from
Eqs.~(\ref{eq:muBEC}--\ref{eq:tBBEC}) and
(\ref{eq:TcBEC}). For comparison, also plotted are the fully numerical
results (black solid lines) from Eqs.~(\ref{eq:neq})-(\ref{eq:PG}). It
is obvious that our asymptotic solution works well for $U/U_c \gtrsim 10$.
Importantly, we see that $T_c$ is governed by $\sqrt{Bt_B}$ in the BEC regime, as in Eq.~(\ref{eq:TcBEC}). A better asymptotic expression for $T_c$, shown as the green dotted line in Fig.~\ref{fig:BEClimit}(d), can be obtained by including higher order effect via fitting $Li_{1/2}^{-2}(1-z)$ with a power law in the range of $z\in [0,0.01]$, which yields
$T_c = 2.22 (nd^2\sqrt{B})^{0.964} t_B^{0.518}$.

It should be noted that the lattice constant $d$ is governed by the
laser wavelength, such as 532 nm or 1064 nm, which usually cannot be
tuned continuously in a typical laboratory. However, one can tune the
density $n$ such that the dimensionless parameter $k_Fd$ can be varied
continuously. As for the parameters $t$ and $U$, normally, one can
tune the ratio of $U/t$ by changing the lattice depth, as described in
Ref.~\cite{Zoller1998PRL}. Since here $t$ is measured in terms of the
Fermi energy $E_F$, therefore, one can tune $t/E_F$ by the density $n$
as well. We note, however, $k_F$ and $E_F$ are locked together via
$E_F=\hbar^2k_F^2/2m$.  Nevertheless, there is already a large room
for tuning these parameters experimentally. More sophisticated
devices, such as a free electron laser \cite{Madey_2016} or a laser
that allows contnuous wavelength tuning \cite{contlaser}, may be used
to allow independent continuous tuning of $d$. In this case, all
parameters can be tuned independently.

\section{Conculsions}
In summary, we study the superfluid transition of a Fermi gas in a
2DOL in the context of BCS-BEC crossover, and pay close attention to
the exotic effect of lattice-continuum mixing.  We find that, for
relatively large $d$ and small $t$ in a 2DOL, a PDW ground state
emerges and $T_c$ vanishes at intermediate pairing strength. In the
BCS and unitary regimes, the nature of the in-plane and overall pairing
may change from particle-like to hole-like, with a negative coefficient
$a_0<0$, and the chemical potential $\mu$ may first increase with
increasing pairing strength.  In the BEC regime,
$T_c \approx \frac{\sqrt{6}ntd^4}{m}|U|^{-1}$, with $\mu \sim -U^2$
and pairing gap $\Delta\sim |\mu|^{3/4} \sim |U|^{3/2}$. These
asymptotic scaling behaviors are qualitatively distinct from their
counterpart in the 1DOL, 3D continuum or 3D lattice cases.  Our
theoretical predictions can be tested in future experiments.

\begin{acknowledgments}
  This work was supported by the NSF of China (Grant No. 11774309
  and No. 11674283), and the NSF of Zhejiang Province of China (Grant
  No. LZ13A040001).
\end{acknowledgments} 

%

\end{document}